\documentclass{ws-procs9x6}

\newcommand{\bra}[1]{\langle {#1} |}
\newcommand{\ket}[1]{| {#1} \rangle}

\begin{document}

\title{Time-dependent density-functional studies on strength
functions in neutron-rich nuclei}

\author{S. Ebata}

\address{Center for Nuclear Study, University of Tokyo,\\
Wako-shi, 351-0198, Japan
%E-mail: an\_author@laboratory.com
}

\author{T. Inakura and T. Nakatsukasa}

\address{RIKEN Nishina Center,
Wako-shi, 351-0198, Japan
%$^*$E-mail: ab\_author@university.com\\
%www.university\_name.edu
}

\begin{abstract}
The electric dipole ($E1$) strength functions have been systematically
 calculated
based on the time-dependent density functional theory (TDDFT),
using the finite amplitude method and the real-time approach to
the TDDFT with pairing correlations.
The low-energy $E1$ strengths in neutron-rich isotopes
show peculiar behaviors, such as sudden enhancement and reduction,
as functions of the neutron numbers.
They seem to be due to the interplay between the neutron shell effect and
the deformation effect.
\end{abstract}


\bodymatter

\section{Introduction}\label{sec:intro}

The electric dipole ($E1$) response is a fundamental mode of excitation
in nuclei and a useful tool to probe the isovector property of nuclei.
The giant dipole resonance may
provide information on the symmetry energy near the saturation density $\rho_0$.
In contrast, the low-energy $E1$ modes, which are often referred to as pygmy
dipole resonances, is sensitive to the nuclear structure and
may carry information on the symmetry energy $E_{\rm sym}(\rho)$
at densities away from $\rho_0$.
Thus, the low-energy dipole modes
in neutron-rich nuclei are of significant interest at present.

The low-energy dipole strengths
have been experimentally observed in several neutron-rich isotopes;
O isotopes \cite{Leistenschneider,Tryggestad}, $^{26}$Ne \cite{Gibelin},
and Sn isotopes \cite{Govaert,Adrich,Ozel,Klimkiewicz,Endres10}.
The low-energy $E1$ strength observed in stable isotopes carries only
less than $\leq 1$ \%\ of the 
Thomas-Reiche-Kuhn (TRK) sum-rule value.
In contrast, it may amount up to about 5 \%\ in neutron-rich nuclei.
Therefore, we may expect a qualitative difference in properties of the
low-energy dipole modes of excitation
between in stable  and in neutron-rich nuclei.

The time-dependent density-functional theory (TDDFT) provides us with a
practical and reasonable description of nuclear strength functions
\cite{Nak12}.
Its rigorous theoretical foundation is given by the one-to-one correspondence
between the time-dependent external potential $v(t)$ and the
time-dependent one-body density $\rho(t)$ \cite{RG84}.
In the Kohn-Sham scheme, it gives the time-dependent Kohn-Sham (TDKS)
equations that are practically identical to the one known as the time-dependent
Hartree-Fock (TDHF) equations in nuclear physics \cite{Neg82}.
An extension for the study of superconducting systems has been also
carried out by including the time-dependent pair potential $\Delta(t)$
and the time-dependent pair density $\kappa(t)$ \cite{WKG94}.
This eventually leads to the
``time-dependent Bogoliubov-de-Gennes-Kohn-Sham (TDBdGKS) scheme'',
which is, in nuclear physics, much more familiar with the name of
``time-dependent Hartree-Fock-Bogoliubov (TDHFB) equation''.
The TDBdGKS equations determine the time evolution of quasiparticle
orbitals whose number is $2M$ where $M$ indicates the dimension
of the single-particle model space.
Since $M$ is significantly larger than the particle number $N$,
it requires a huge computational task and becomes a challenging subject in
computational nuclear physics \cite{SBMR11,Nak12}.

In this paper, alternative approaches based on the TDDFT,
which are computationally more feasible than the full solution of the
TDBdGKS equations,
are utilized
to study the properties of the low-energy $E1$ strength.
These methods will be briefly explained in Sec. \ref{sec:theory}.
The numerical applications are given in Sec. \ref{sec:E1}.

\section{Theoretical tools}\label{sec:theory}

We use the canonical-basis formulation of the TDBdGKS method and
the finite amplitude method (FAM)
to obtain low-energy $E1$ strength functions in neutron-rich nuclei.
We briefly recapitulate these methods in the followings.
Details of the method can be found in references given below.

\subsection{Canonical-basis TDBdGKS equations}

The real-time calculations of the TDKS equations have been carried out
in real space \cite{Neg82,NY01,NY05}.
However, the computational cost is significantly increased by inclusion of
the pairing correlation (TDKS $\rightarrow$ TDBdGKS), which makes
practical applications very difficult.
This numerical cost can be reduced, by several orders of magnitude,
introducing an approximation for the time-dependent pair potential,
similar to the BCS approximation in static cases \cite{Eba10}.
This may lead to the following set of equations:
% which we call
%``Canonical-basis TDHFB'' (Cb-TDHFB) equations:
\begin{eqnarray}
\label{td_cs_1}
i\frac{\partial}{\partial t} \ket{\phi_k(t)} 
&=& \left( h(t) - \eta_k(t) \right) \ket{\phi_k(t)} , \\
\label{td_cs_2}
i\frac{\partial}{\partial t} \ket{\phi_{\bar k}(t)} 
&=& \left( h(t) - \eta_{\bar k}(t) \right) \ket{\phi_{\bar k}(t)} , \\
\label{td_rho}
i\frac{\partial}{\partial t} \rho_k(t)
&=& \kappa_k(t)\Delta_k^*(t) - \kappa_k^*(t) \Delta_k(t) , \\
\label{td_kappa}
i\frac{\partial}{\partial t} \kappa_k(t)
&=& \left( \eta_k(t) + \eta_{\bar k}(t) \right) \kappa_k(t)
+ \Delta_k(t) \left( 2\rho_k(t) -1 \right) .
\end{eqnarray}
Here, 
Eqs. (\ref{td_cs_1}) and (\ref{td_cs_2}) describes the time
evolution of a pair of canonical states, $\ket{\phi_k(t)}$ and
$\ket{\phi_{\bar k}(t)}$.
Their occupation and pair probabilities are given by
Eqs. (\ref{td_rho}) and (\ref{td_kappa}).
The Hamiltonian $h(t)$ is a functional of density $h[\rho(t)]$,
and $\eta_k(t)$ and $\eta_{\bar k}(t)$ are arbitrary real parameters.
In numerical calculations in Sec.~\ref{sec:E1}, we adopt
$\eta_k(t)=\bra{\phi_k(t)}h(t)\ket{\phi_k(t)}$ 
$\eta_{\bar k}(t)=\bra{\phi_{\bar k}(t)}h(t)\ket{\phi_{\bar k}(t)}$.
The computational task for solution of these equations is roughly
similar to that of TDKS equation.

\subsection{Finite amplitude method for linear response}

The finite amplitude method (FAM) \cite{NIY07} is another feasible
approach to the linear response calculation.
The formulation has been extended to superfluid systems as well\cite{AN11}.
The method allows us to easily construct a computer code for the
linear response calculation based on the TDDFT.
The FAM has been applied to the coordinate-space
representation \cite{NIY07,INY09,INY11,AN11,AN13}, and
to the harmonic-oscillator-basis representation \cite{Sto11}.
The essential idea is that the complicated residual induced part of the
Hamiltonian $\delta h$ can be
calculated in terms of the finite difference associated with the
non-Hermitian density $\rho_\eta$.
Provided that $\ket{\phi_k}$ are the canonical single-particle states
at the ground state ($h_0 \ket{\phi_k}=\epsilon_k \ket{\phi_k}$ and
$\rho_0=\sum_{i:\textrm{hole}}\ket{\phi_i}\bra{\phi_i}$),
\begin{eqnarray}
\label{delta_h}
\delta h &=& \frac{1}{\eta} \left( h[\rho_\eta] - h_0 \right) ,\\
\label{rho_eta}
\rho_\eta &=& \rho_0+\eta\delta\rho 
=\sum_{i:\textrm{hole}} \ket{\psi_i}\bra{\bar{\psi}_i} ,
\end{eqnarray}
where
\begin{eqnarray}
\ket{\psi_i} &=& \ket{\phi_i} + \eta \sum_{m:\textrm{particle}} X_{mi}\ket{\phi_m} , \\
\bra{\bar{\psi}_i} &=& \bra{\phi_i} + \eta \sum_{m:\textrm{particle}} Y_{mi}\bra{\phi_m} .
\end{eqnarray}
In the FAM,
the induced field $\delta h$ for a given $(X,Y)$
can be estimated using Eq. (\ref{delta_h}).
This only requires us to calculate the single-particle (Kohn-Sham)
Hamiltonian $h[\rho]$ with different bra's and ket's in Eq. (\ref{rho_eta}).
Then, we resort either to iterative algorithms \cite{NIY07,INY09,Sto11}
or to the matrix diagonalization \cite{AN13}, in order to obtain
solutions of the linear-response equation $(X,Y)$.

\section{Low-energy $E1$ strength}\label{sec:E1}

The $E1$ strength functions in even-even isotopes are
systematically calculated.
For relatively light nuclei,
the FAM was used neglecting the pairing correlations \cite{INY11}.
For heavier nuclei, in which the pairing is expected to play an important
role, we use the real-time method based on the
canonical-basis TDBdGKS method with a time-dependent perturbation
of the external $E1$ field\cite{NY05,Eba10}.

\subsection{Shell effects}

\begin{figure}[tb]
\begin{center}
\includegraphics[width=0.6\textwidth,keepaspectratio]{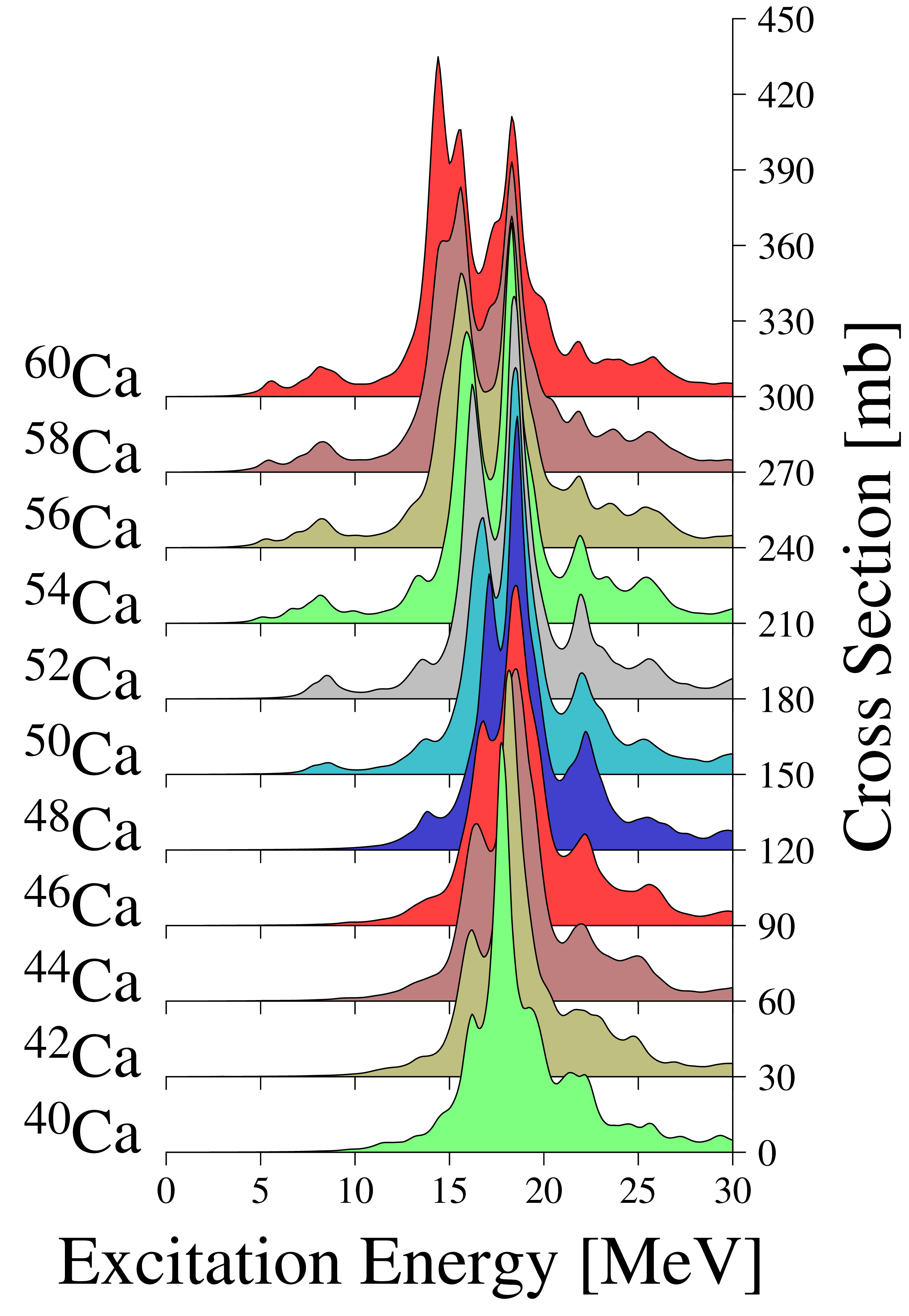}
\caption{Photoabsorption cross sections in Ca isotopes,
calculated using the FAM with the SkM* functional.
The origin of the vertical axis is shifted upward by 30 mb as the
neutron number increases by two.
}
\label{fig:Ca}
\end{center}
\end{figure}

The photoabsorption cross sections for Ca isotopes
($N=20-40$),
estimated in the dipole approximation,
are shown in Fig.~\ref{fig:Ca}.
For Ca isotopes with $N=20-28$, we see only negligible $E1$ strengths
below $E=10$ MeV.
In contrast, those with $N=30-40$ shows sizable $E1$ strengths below
10 MeV.
This indicates that a sudden appearance of the low-energy $E1$ modes takes
place beyond the magic number $N=28$ in Ca isotopes.
We have confirmed the same behavior in neighboring isotopes \cite{INY11}.

Similar jumps of the low-energy $E1$ strengths can be
observed at $N=14\rightarrow 16$ and $N=50\rightarrow 52$ \cite{INY11}.
These indicate a strong shell effect on the appearance of
the low-energy $E1$ strength.
These numbers are related to the occupation of single-particle orbitals
with low orbital angular momenta, such as $s_{1/2}$ ($N=14\rightarrow 16$),
$p_{3/2}$ ($N=28\rightarrow 30$), and
$d_{5/2}$ ($N=50\rightarrow 52$).
When these low-$\ell$ orbitals are weakly bound
having spatially extended characters,
we may expect the threshold effect which may enhance the $E1$ strengths
near the neutron emission threshold energy.
Thus, we suppose that the low-energy $E1$ strengths predominantly possess
a single-particle nature.

\subsection{Effects of deformation and separation energy}

\begin{figure}[tb]
\begin{center}
\includegraphics[width=0.5\textwidth,angle=-90]{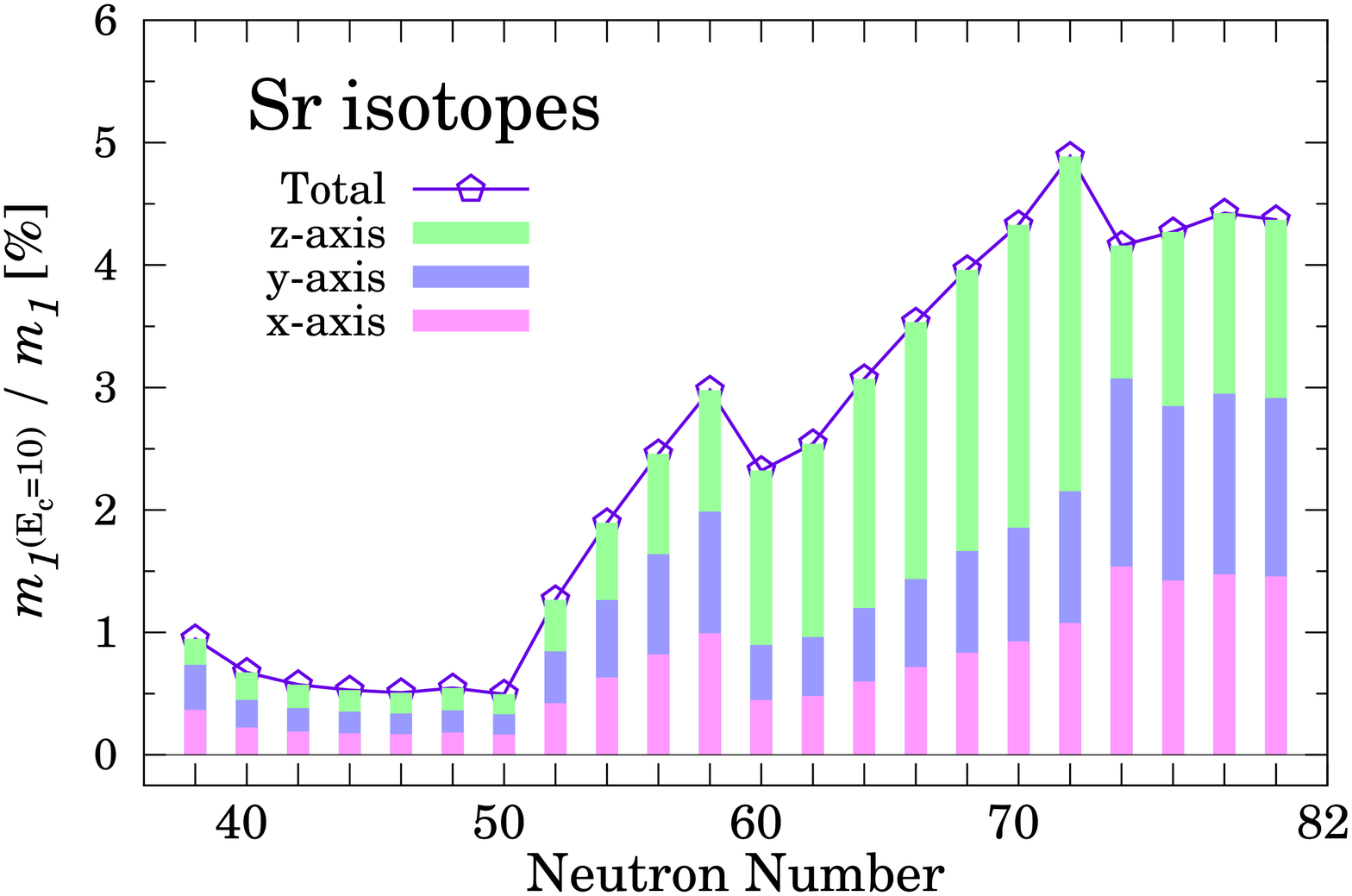}
\caption{$m_1(E)/m_1(\infty)$ with $E=10$ MeV for even-even Sr isotopes.
The calculation was performed with the SkM* functional using the
canonical-basis TDBdGKS equations.
See text for details.
}
\label{fig:Sr}
\end{center}
\end{figure}

Let us move toward heavier isotopes.
We use the canonical-basis TDBdGKS method to calculate the $E1$ strength
distribution \cite{Eba10,Nak12}.
We have found that the effect of pairing correlation is not so significant,
in general.
However, for selected nuclei, the pairing correlation affects
the shape of the ground state, which modifies the $E1$ properties accordingly.

We calculate the integrated energy-weighted $E1$ strength up to $E=E_c$,
defined as
\begin{equation}
\label{m1}
m_1(E_c)\equiv \int_0^{E_c} E \frac{dB(E1)}{dE} dE
= \sum_{\mu=x,y,z} \int_0^{E_c} E \frac{dB(E1;\mu)}{dE} dE .
\end{equation}
In Fig. \ref{fig:Sr}, the ratio of $m_1(10\textrm{ MeV})$
to $m_1(\infty)$ is shown for Sr isotopes.
The ratio is less than 1 \%\ for isotopes with $N\leq 50$.
Then, the ratio jumps up beyond $N=50$,
which is consistent with the argument given above.
However, there are sudden drops of the low-energy $E1$ ratio at
$N=58\rightarrow 60$ and
$N=72\rightarrow 74$.
This seems to be due to changes of the ground-state deformation.

The calculated ground states in even-even Sr isotopes with $N=40-50$ are
all spherical ($\beta=0$).
The two-neutron separation energies, which are equal to twice of the
chemical potential, gradually decrease as the neutron number increases.
Then, the calculation predicts that
the onset of the ground-state deformation takes place
at $N=58\rightarrow 60$, from spherical to prolate shapes
($\beta=0\rightarrow 0.37$).
This shape transition result in the increase of the 
two-neutron separation energy and
the decrease of the low-energy $E1$ strengths.
The deformation stays rather constant for $N=60-72$, with prolate shapes
of $\beta=0.34-0.38$ and with
decreasing separation energies as increasing the neutron number.
At $N=72\rightarrow 74$, again, the shape transition takes place,
from prolate to oblate shapes,
$(\beta,\gamma)=(0.34,0)\rightarrow (0.14,60^\circ)$.
This shape change accompanies the increase of the separation energy
and decrease of the low-energy $E1$ strength.

The low-energy strength can be decomposed into
those associated with $x$, $y$, and $z$ directions,
as in the last equation in Eq. (\ref{m1}).
This decomposition is also shown in Fig. \ref{fig:Sr}.
The $z$ axis is chosen as the symmetry axis for axially deformed
nuclei.
For prolate Sr isotopes with $N=60-72$,
the strength associated with the $z$ ($K=0$) component is dominant.
The $K=0$ dominance was also reported for neutron-rich Sn isotopes
using the relativistic quasiparticle random phase approximation \cite{PKR09}.
This was interpreted by the conjecture that the neutron skin is thicker
in the $z$ direction than the $x$ and $y$ directions.
We calculate the neutron-skin thickness in the $x$ direction as 
$\Delta x_{np}=\sqrt{\langle x^2\rangle_n} -\sqrt{\langle x^2\rangle_p}$,
and those for $y$ and $z$ directions in exactly the same way.
It turns out that $\Delta r_{np}$ is even larger with respect to
the $x$ ($y$) direction than the $z$ direction, for prolate nuclei.
Therefore, the observed $K=0$ dominance in the low-energy $E1$ strength
cannot be attributed to the different neutron skin thickness.
This suggests that these low-energy strengths are not associated with
the skin modes.

\section{Summary}

The $E1$ strength functions have been systematically calculated
with the finite amplitude method and the real-time method,
based on the time-dependent density functional theory.
The low-energy $E1$ strength distributions in stable and neutron-rich isotopes
were estimated from these calculations.
We have found a strong neutron shell effect and have identified
magic numbers for the appearance of low-energy $E1$ modes.
The deformation and separation energies
also play an important role in the low-energy $E1$ strength distribution.

\section{Acknowledgments}

This work is supported by MEXT/JSPS KAKENHI Grant numbers
20105003 and 21340073.
The numerical calculations were partially performed on
T2K at Center for Computational Sciences, University of Tsukuba,
Hitachi SR16000 at YITP, Kyoto University,
and the RIKEN Integrated Cluster of Clusters (RICC).

\end{document}